# Electric field dependence of Raman-active modes in single-wall carbon nanotube thin films

Giovanni Fanchini, Husnu Emrah Unalan and Manish Chhowalla

Materials Science and Engineering – Rutgers University – Piscataway, NJ08854, USA

We report on electrical Raman measurements in transparent and conducting single-wall carbon nanotube (SWNT) thin films. Application of external electric field results in downshifts of the D and G modes and in reduction of their intensity. The intensities of the radial breathing modes increase with electric field in metallic SWNTs, while decreasing in semiconducting SWNTs. A model explaining the phenomenon in terms of both direct and indirect (Joule heating) effects of the field is proposed. Our work rules out the elimination of large amounts of metallic SWNTs in thin film transistors using high field pulses. Our results support the existence of Kohn anomalies in the Raman-active optical branches of metallic graphitic materials [Phys. Rev. Lett. 93 (2004) 185503].

Transparent and conducting single-wall carbon nanotube (SWNT) thin films are two dimensional, low density networks of SWNTs [1] which are interesting both fundamentally and technologically. Recently, it has been noticed [2] that the Drude relaxation times of SWNT thin films, and therefore their optolectronic properties, may be controlled by inter-tube processes. This could represent an important difference with respect to individual SWNTs where intra-tube processes dominate [3]. Technologically, the ability [1] to tailor the optical absorption coefficient and conductivity of SWNT thin films over several orders of magnitude makes them attractive for transparent and flexible electronics [4-7]. During SWNT thin film transistor fabrication, it is a common practice to pre-condition the SWNT network using high voltage pulses to improve the on/off ratio through supposed preferential elimination of metallic SWNTs by Joule heating [4-7]. This effect is claimed on the basis of the decrease in channel conductivity and a similar effect occurring in individual SWNTs, but little information on the modifications of the SWNTs in thin films after voltage application is available and a Raman study is lacking.

In this Letter, we report on the Raman measurements of SWNT thin films recorded under external voltages. We have found that, although the conductivity strongly decreases, the changes in the Raman peaks are to the largest extent reversible. The films were deposited on glass using the method of Wu et al [8] from 10, 30, and 50 mL of a 2 mg/L suspension of purified HiPCO SWNTs [2,9]. Gold electrodes (width 1 mm, distances 20 and 60 μm) were defined on each substrate. External voltages of 0-15 V (leading to electric fields $E_{ext}$ = 0-7500 V $cm^{-1}$) were applied during the Raman measurements using a GW GPS-1850D power supply. The spectra were recorded in air on a Renishaw InVia spectrometer. Our setup for electrical Raman measurements is presented in Fig. 1a. Low laser powers (12.5 $\mu W/\mu m^2$ at 1.96 eV excitation, 25 $\mu W/\mu m^2$ at 1.57 eV) were used and tested to not produce laser heating. The current was simultaneously monitored using a Keithley 195A multimeter. Each series of Raman spectra at the varying voltage was recorded on the same spot in order to attain comparable signal intensity. After measuring at any given external field, sufficiently low fields (500 V/cm, leading to undetectable changes to the Raman signal) were applied so that a low-field Raman spectrum and the sample conductance could be recorded simultaneously after each measurement under external field.

The typical variation of the G-bands [3] and the doubly-resonant D-band [10] under the influence of an electric field are shown in Fig.1b and c, respectively. A clear decrease in the peak position ($\Omega$) and intensity ($I_S$) for both bands with increasing electric field can be observed. Similar shifts in the Raman bands under the influence of electric fields has been observed in ferroelectrics [11]. In contrast, electrochemical Raman measurements of SWNT electrodes in aqueous environments [12] resulted in upshifts of the G-bands, which clearly points to differences between our and such experiments. It is tempting to assign the observed downshifts to a voltage-induced increase in phonon temperature (T) of the SWNTs. However, in the absence of other effects, the intensity of the Stokes Raman peaks should increase with temperature [13] according to:

$$I_S \sim n(\hbar\Omega/k_BT) + 1 \qquad (1)$$

(where n is the Bose occupation number). Furthermore, the linewidths of the G peaks are expected to be strongly broadened at high temperature [14], which is not the case in our spectra, whose linewidths are almost independent of the external field. Thus, our experiment cannot be satisfactorily explained in terms of strong temperature increase.

Another interpretation, which we will dismiss below, might deal with electromechanical strain owing to the strong elasticity of SWNTs [15] which could also result in a reversible downshift of the G-peak [16].

It is therefore critical to investigate whether the Raman effects and the decrease in conductivity are reversible or irreversible. Fig.2a-b show the Raman peaks recorded immediately after releasing each external field used for the measurements shown in Fig.1b-c. The recovery of the G and D peaks to their original frequencies is evident. Peak intensities not only recover but they also increase slightly compared to their pristine value before field application. From the data in Fig.2a-b we conclude that the observed effects on the G and D peaks are indeed reversible. The data from Fig.1b and 2a are summarized in Fig.2c, showing the ratios between the intensity of the G-bands before and after the external field release for the samples and excitation energies investigated in this study.

While the Raman effects in our experiments are reversible, the conductance does not recover at all after the external field release (Fig.2d). Similar decrease in conductance was found to strongly improve the on/off ratio of thin film transistors and claimed to be due to burning of the metallic SWNTs [4-7]. In our study, the reversibility in decrease of the peak intensities rules out the burning of large amount of SWNTs. We corroborate this idea by analyzing the radial breathing mode (RBM) of individual SWNTs in our films, as shown below. The absence of a reversible decrease in conductance (shown in Fig. 2d for the 50 mL sample) between the measurements performed during high field application (open circles) and after release of the electric field (solid circles) is critical since it proves that the observed downshift of the Raman bands cannot due to electromechanical strain. Strain would indeed provide piezoresistivity, implying that conductance would recover after strain release, as found for individual SWNTs suspended in air or on tensile membranes [17]. We attribute the lack of electrical strain in our experiments to the strong adhesion of our films to a stiff substrate. However electromechanical effects might be relevant in electrical Raman experiments on suspended SWNTs [17]. We also detected identical Raman effects on SWNT films embedded in polyethylene-imine, an insulating oxygen-repellent polymer able to switch SWNT thin film transistors from p- to n-type [5]. In this case the decrease in conductance was still permanent, but lower.

In order to provide insight in to the influence of the applied field $E_{ext}$ on the Raman bands, we will discuss our results in terms of strong field-dependent fluctuations in the dielectric response of SWNT thin films. This is a similar model to what has been proposed in ferroelectrics [11]. We will assign, in SWNT thin films, such fluctuations to the presence of Kohn anomalies in the D and G phonon branches.

Let us first recall that, in percolating SWNT networks, the electronic confinement is released and the wave-functions extend over several SWNTs, both semiconducting (s-SWNTs) and metallic (m-SWNTs) and $\tau$, the Drude relaxation times, depend on the network density and not on intrinsic properties of SWNTs [2]. Relaxation indicates that the external field displaces the Fermi sphere through a shift in momentum [18]

$$\hbar \cdot \Delta k = e \cdot \tau \cdot E_{ext} \qquad (2)$$

Let us then recall that the strong Raman activity of the G and D bands of metallic SWNTs should correspond to exceptionally strong electron-phonon coupling [3,10,19], whose origin had been unclear for a long time. Recently, Piscanec et al [20] offered an explanation for such important phenomena through demonstration of the existence of Kohn anomalies in the screening of ions in metallic graphitic materials. In general, Kohn anomalies occur when the size of the Fermi surface is comparable to the phonon wavevector **q** [20-22]. As discussed by Piscanec et al [20], the π and π* bands in graphite and m-SWNTs touch the Fermi level ($E_F$) at the **K**-point which results in a very small Fermi wavevector ($\mathbf{k_F} \sim \mathbf{0}$) whose modulus approaches those of the G and D phonon wavevectors ($\mathbf{q} = \Gamma = \mathbf{0}$ and $\mathbf{q} = \mathbf{q'} - \mathbf{K} = \mathbf{0}$). Since $|2\mathbf{k_F}|^{-1}$ represents the typical scale length for screening a point-like disturbance [22], the condition $\mathbf{q} \sim 2\mathbf{k_F} \sim \mathbf{0}$ requires that infinite distance is needed for the electrons to fully screen an optical phonon.

It is well known that Kohn anomalies in 1-D solids lead to logarithmic divergence of the static dielectric response, $\varepsilon(q{\sim}2k_F, \hbar\omega{\sim}0)$, while in 3-D solids the divergence only affects the first derivative of this quantity. Dealing with 1-D electronic structures, it is then obvious that little fluctuations in the electron momenta (e.g. by applying a constant external field) correspond to strong, non-negligible, fluctuations in the dielectric response [23]. We shall treat such fluctuations in the framework of the Lindhard model [24]. Accordingly, the dynamic dielectric response of the anomalously screening electrons in the presence of a change in momentum $\hbar \cdot \Delta k(E_{ext})$ is given by [22]

$$\varepsilon[\Delta k(E_{ext}),\hbar\omega] - 1 = -4e^2\varepsilon_0 \lim_{\substack{k_F \to 0 \\ q/2k_F \to 1}} \int dk \frac{f(k_F + \Delta k + q/2, T) - f(k_F + \Delta k - q/2, T)}{\in_{k+\Delta k+q/2} - \in_{k+\Delta k-q/2} - \hbar\omega} \qquad (3)$$

where the Fermi-Dirac population probability f(k,T) will be taken to be approximately linear in the energy domain $E_F \pm k_BT/2$ and 1 or 0 elsewhere. Since the dielectric responses obtained from ellipsometry at our Raman excitation energies ($\hbar\omega$ = 1.57-1.92 eV) follow a Drude behavior [2], $\in_k$ will be taken to be the dispersion relation for free electrons ($\in_k = \hbar^2k^2/2m$). Care should be taken in evaluating eq. (3), because very slight changes in q, ω and T can lead to fluctuations of $\varepsilon(q, \hbar\omega)$ from 1 to infinity [23]. Therefore, since a very small value of $k_F$ is expected in our percolating networks of metallic SWNTs, we will estimate eq. (3) at $k_F$ tending to zero with the same zero-th order of q, thus leading to a finite ratio $q/2k_F \to 1$. Thus, straightforward evaluation of eq. (3) gives $\varepsilon(0, \hbar\omega > 0) - 1 = 0$ in the absence of external electric field, while in the presence of field

$$\varepsilon(\Delta k,\hbar\omega)-1=\frac{8e^2}{\varepsilon_0 k_B T}\left(\sqrt{\frac{mk_B T}{\hbar^2}+\Delta k^2}-\sqrt{\frac{mk_B T}{\hbar^2}-\Delta k^2}\right) \quad \mathit{for}\; k_B T \gg \frac{\hbar^2\Delta k^2}{m}. \quad (4)$$

The intensity and the frequency of the Raman-active optical phonons are related, via the electron-phonon coupling, to the dynamic dielectric response $\varepsilon(q\sim 2k_F, \hbar\omega)$ [22]. Therefore, knowledge on the dynamic dielectric response will now allow us to extract information on the G and D mode frequencies. Especially, if the screening is assumed to (perturbed by temperature and electric field) largely determine frequencies of the screened optical phonons, then the Raman shifts in the presence $[\Omega(E_{ext})]$ and absence $[\Omega(0)]$ of field can be related by [11,25]:

$$\Omega_T(E_{ext})^2 \cdot \varepsilon(\Delta k(E_{ext}),\hbar\omega) = \Omega_T(0)^2 \cdot \varepsilon(0,\hbar\omega) \quad (5)$$

The anharmonic modifications of the SWNT structure can be included by assuming a temperature dependent zero-field Raman shift $\Omega_T(0) \approx \Omega_{300K}(0) - X_T \cdot T$ ($X_T \approx 0.01$ cm$^{-1}$K$^{-1}$, for the G-peak [26]). However, we anticipate that they would not significantly affect the parameters achievable by fitting our model with the experiment (relaxation times and temperatures change below 30-40%). Replacement of $\varepsilon(\Delta k(E_{ext}),\hbar\omega)$ and $\varepsilon(0,\hbar\omega)$ from eqs. (3-4) into eq.(5) leads to the following relation for the frequency of the Raman optical modes upon temperature and external field increase, as plotted in Fig.3a:

$$\Omega_T(E_{ext}) = \Omega_T(0)\cdot\left[1+\frac{8e^2}{\hbar\varepsilon_0 k_B T}\left(\sqrt{mk_B T+\frac{e^2\tau^2}{2}E_{ext}^2}-\sqrt{mk_B T-\frac{e^2\tau^2}{2}E_{ext}^2}\right)\right]^{-1/2} \quad (6)$$

A comparison between our model and experimental results is shown in Fig.3b-c. Fits were obtained using Drude relaxation times of $\tau \sim 10^{-15}$ s and assuming the temperature of the Raman-active phonons rising linearly from $T_{min}$ = 300K to $T_{max}$ = 500-700K at external fields $E_{ext}$ = 0-7500 V/cm [27]. Note that these temperatures are far too low to burn the m-SWNTs. In the framework of our model, the decrease in intensity of the Raman modes can be easily explained since the Stokes/Anti-Stokes Raman cross sections $I_{S/AS} \sim |\partial\varepsilon/\partial u_{//}|^2 + |\partial\varepsilon/\partial u_{\perp}|^2$ [28] decrease at increasing fields in direction longitudinal to the field, while remaining unchanged in transversal directions [29,30].

We have also examined the RBMs of our thin films in order to investigate the influence of the electric field on (n,m)-SWNTs with various chiralities. The RBMs are shown in Fig.4a-b. It can be seen from Fig.4a that at $\hbar\omega$ = 1.96 eV, where both s- and m-SWNTs are excited, the intensities of the RBMs of s-SWNTs decrease with increasing electric field while the intensities of the RBMs of m-SWNTs increase. In Fig.4b ($\hbar\omega$ = 1.57 eV) where only s-SWNTs are sampled, the intensities of the RBMs decrease with increasing electric field. In contrast, after the electric field has been released, it can be seen in Fig.4c-d that the RBM intensity always slightly increases in both s- and m-SWNTs. The most interesting feature of this slight increase is that it remains permanent subsequent to the field release. Thus, the decrease in measured conductance shown in Fig.2d cannot be correlated to the claimed preferential elimination of m-SWNTs.

The increase in intensities of the RBMs of m-SWNTs is an expected effect if the temperature increase of the optical phonons is the determining factor of peak intensity, according to eq.(1). The decrease in the intensities of RBMs in s-SWNTs is not consistent with Joule heating as the determining factor in controlling the intensity of the RBMs.

Such a decrease may be due to direct effects of the electric field, as described above for the D and G modes. It is not surprising that in a SWNT network the effects of the field may extend to s-SWNTs which do not directly involve Kohn anomalies. Indeed, since our films are percolating [9], there is no electron confinement in either m- or s-SWNTs.

In conclusion, we reported on the changes in Raman peaks of SWNT thin films as a function of an external electric field. We assign such effects to anomalous electron-phonon interactions. Anharmonic (phonon-phonon) effects such as thermal dilatation or electromechanical strain seem to be less important. We dismiss the idea that, in SWNT thin films, voltage pulses burn large amounts of m-SWNTs. Rather, thermal oxidation [32] or selective cutting of the m-SWNTs, or the elimination of very little amounts of m-SWNTs on some critical percolative pathways, may reduce the 'off' currents, thus improving the transistor performance. Finally, electrical Raman spectroscopy will be a new and powerful analytic technique for characterizing thin films and devices incorporating one-dimensional nanostructures.

## References

[1] L. Hu, et al, Nano Lett. 4 (2004) 2513

[2] G.Fanchini, et al, Appl. Phys. Lett. 89 (2006) 191919

[3] M.D. Dresselhaus, P.C. Eklund, Adv. Phys. 49 (2000) 705.

[4] E.S. Snow, et al, Appl. Phys. Lett. 82 (2003) 2145

[5] Y. Zhou, et al, Nano Lett. 4 (2004) 2031.

[6] R. Seidel, et al, Nano Lett. 4 (2004) 831

[7] T. Takenobu, et al, Appl. Phys. Lett. 88 (2006) 033511.

[8] Z.Wu, et al, Science 305 (2004) 1273

[9] H.E. Unalan, et al, Nano Lett. 6 (2006) 2513

[10] H. Thomsen, S. Reich, Phys. Rev. Lett. 85 (2000) 5214

[11] J.M. Worlock, P.A. Fleury, Phys. Rev. Lett. 19 (1967) 1176

[12] P. Corio, et al, Chem. Phys. Lett. 370 (2003) 675; 392 (2004) 396; K. Okazaki, et al, Phys. Rev. B 68 (2003) 035434

[13] M.Cardona in Light Scattering in Solids (M.Cardona ed) Springer, Berlin, 1975, p11

[14] A. Jorio, et al, Phys. Rev. B 66 (2002) 115411; M.N. Iliev, et al, Chem. Phys. Lett. 316 (2000) 217

[15] V.N. Popov, et al, Phys. Rev. B 61 (2000) 3078; D. Sanchez-Portal, et al, Phys. Rev. B 59 (1999) 12678

[16] S.B. Cronin, et al, Phys. Rev. Lett. 93 (2004) 167401.

[17] E.D. Minot, et al, Phys. Rev. Lett. 90 (2003) 156401; J. Cao, et al, Phys. Rev. Lett. 90 (2003) 157601; R.J.Grow, et al, Appl. Phys. Lett. 86 (2005) 093401.

[18] M. Dressel, G.Gruner, Electrodynamics of Solids, Cambridge University Press, 2002

[19] A.C. Ferrari, J. Robertson, Phys. Rev. B 61 (2000) 14095

[20] S. Piscanec, et al, Phys. Rev. Lett. 93 (2004) 185503

[21] V.N. Popov, P. Lambin, Phys. Rev. B 73 (2006) 085507.

[22] N.W. Ashcroft, N.D. Mermin, Solid State Physics, Saunders College, Philadelphia, 1976, p515

[23] For instance, at T = 0K, eq. (3) would lead to $\varepsilon(\Delta k, \hbar\omega)$-1 ~ $\log\{[(x+1)^2-(x_E+y)^2] / [(x-1)^2-(x_E+y)^2]\}$, with $x=q/2k_F$, $x_E=\Delta k(E_{ext})/2k_F$ and $y=m\omega/\hbar q$ so that Kohn anomalies (x = 1) correspond, in the absence of field, to an infinite static response $\varepsilon(0,0)-1 \rightarrow \infty$ and a null dynamic response $\varepsilon(0, \hbar\omega)-1 = 0$ while, in the presence of field, $\varepsilon(\Delta k,0)-1 \rightarrow \infty$.

[24] It has also been recently suggested that the electron-phonon interaction in individual metallic SWNTs is strong enough that it cannot be treated within the Born-Oppenheimer approximation (A.C. Ferrari http://kypros.physics.uoc.gr/smac_2006/Ferrari_abs.pdf) In this case the use of the Lindhard model would be inadequate. However in SWNT networks, unlike in individual SWNTs, we may expect that such dynamic effects are increasingly quenched since each metallic tube is interconnected by many semi-conducting tubes, which are not directly affected by Kohn anomalies.

[25] N.W. Ashcroft, N.D. Mermin, Ref. [22] p513-523

[26] T. Uchida, et al, Chem Phys Lett 400 (2004) 341; Z. Zhou, et al, J Phys Chem B 110 (2006) 1206. Nevertheless, while the changes achieved including possible anharmonic effects in our model are low, the influence on the electron screening of the alternating electric fields used for Raman excitation might also be considerable. This would suggest that our model might also be useful in explaining the temperature dependence of the

Raman peaks of SWNT thin films in the absence of a constant external field.

[27] Note that the Drude relaxation times determined within our model are of the order of magnitude available in literature (see e.g. [22] p10) and they correspond to the ones achievable by ellipsometry from our samples. However, these values might also have been affected by defects or impurities present in HiPCO SWNTs.

[28] $u_{//}$ and $u_{\perp}$ represent, in each SWNT, the modes polarization longitudinal and transversal to the electric field.

[29] A.S. Barker, R. Loudon, Rev. Mod. Phys. 44 (1972) 18.

[30] G. Fanchini, et al, J. Appl. Phys. 91 (2002) 1155

[31] H. Telg, et al, Phys. Rev. Lett. 93 (2004) 177401

[32] S-H. Jhi, et al, Phys. Rev. Lett. 85 (2000) 1710

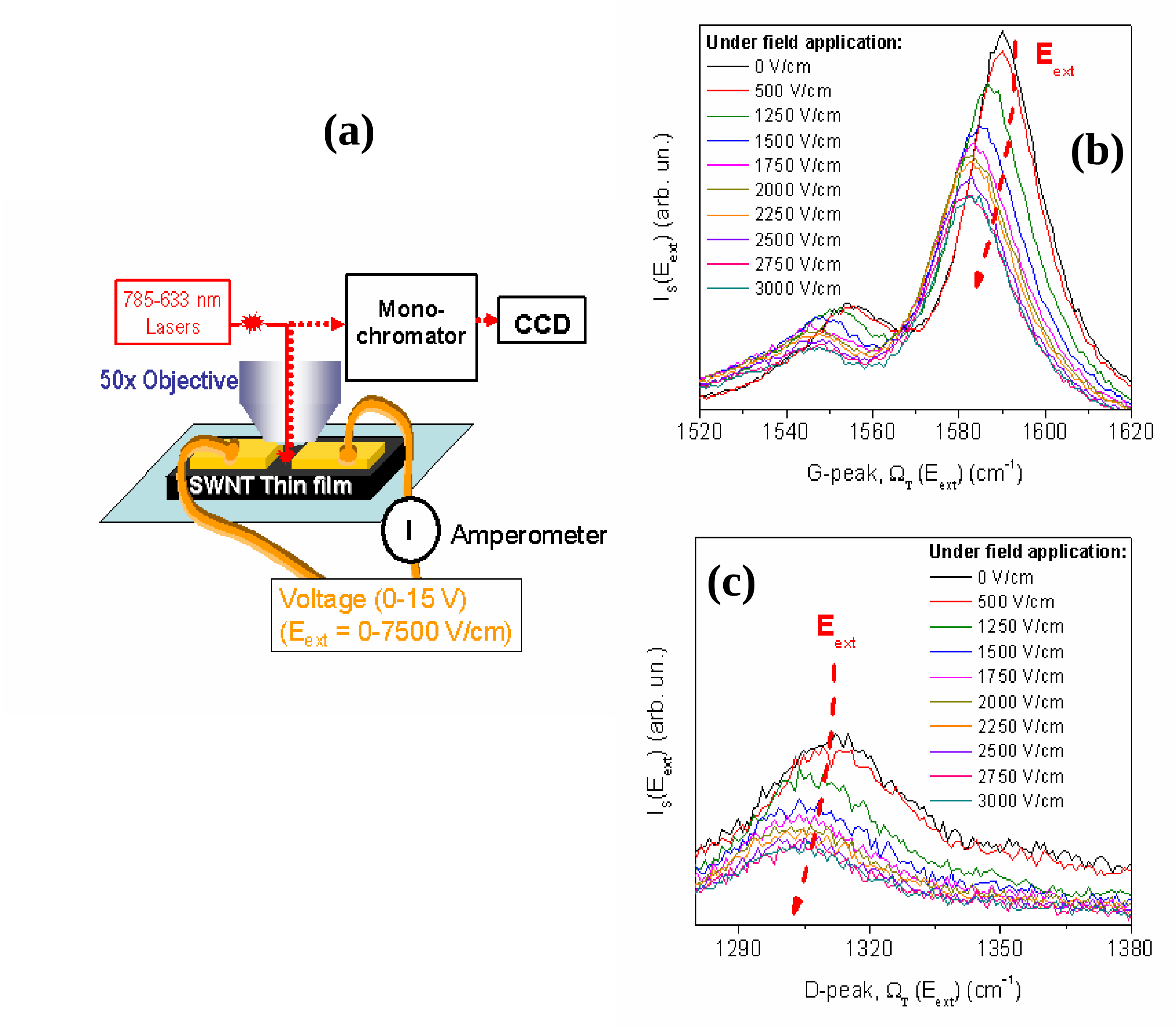

**FIG. 1**

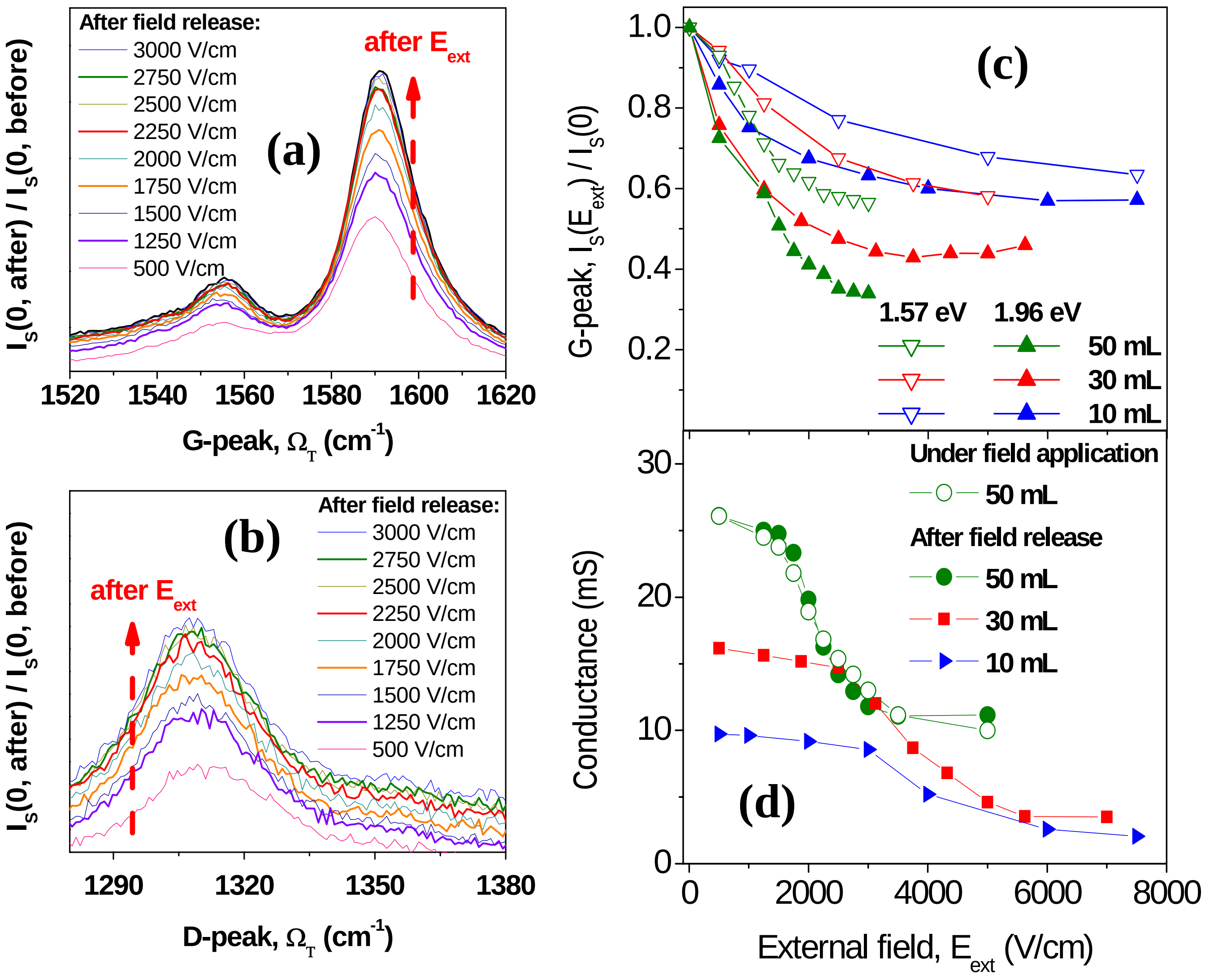

FIG. 2

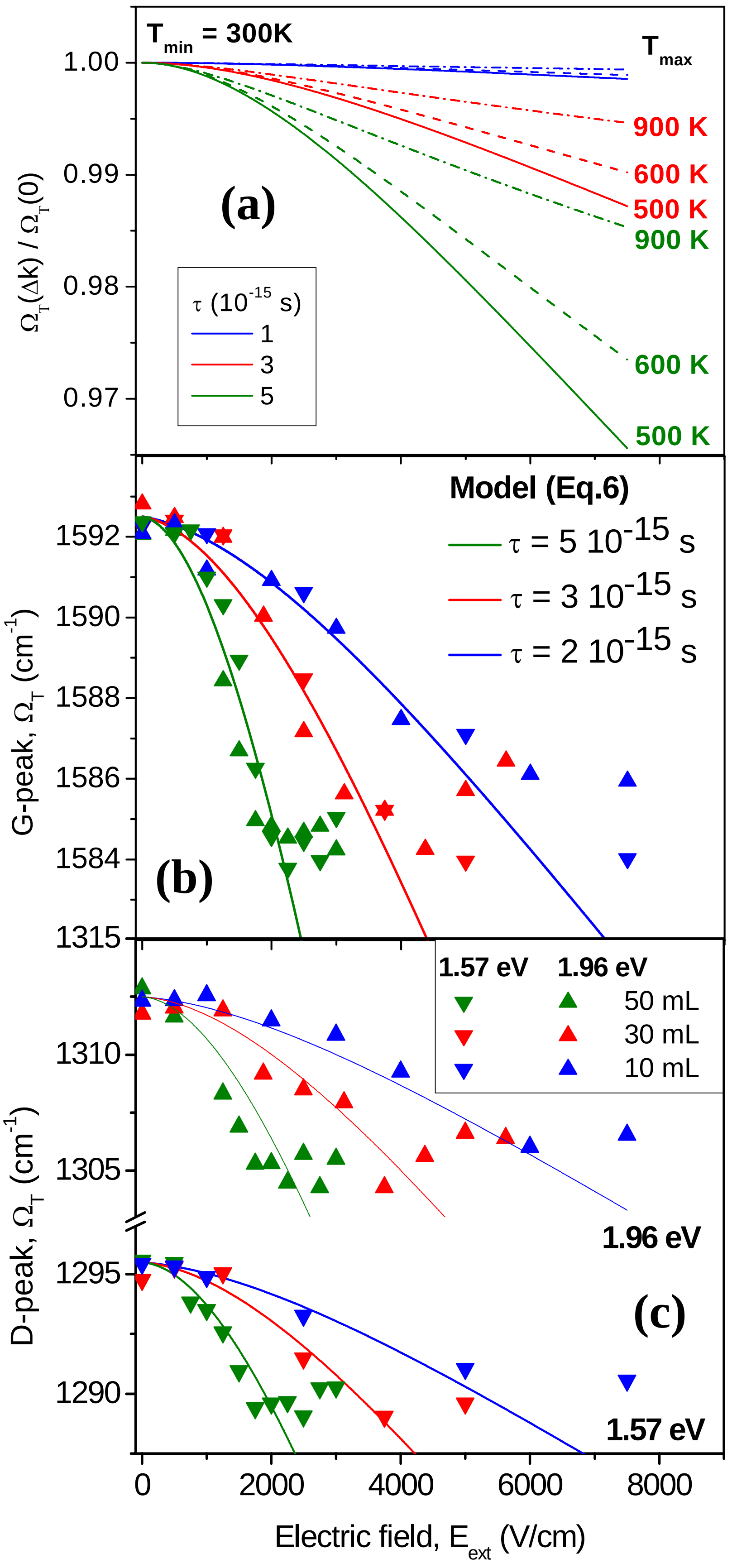

**FIG. 3**

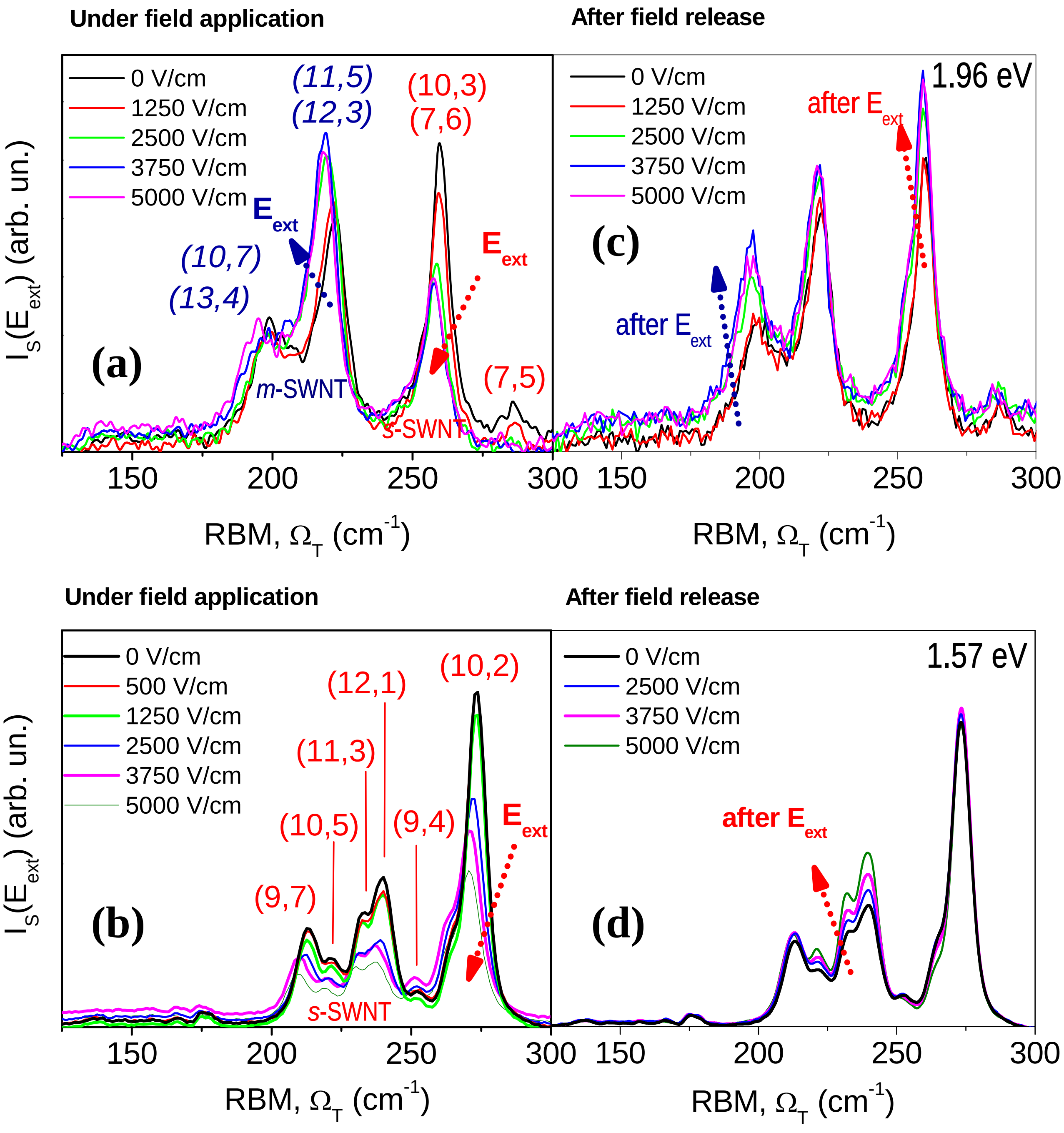

**FIG.4**

## Figure Captions

**Fig.1** – (a) Schematic of the setup for electrical Raman measurements. (b) Typical dependence of the Raman modes on the electric field ($E_{ext}$) for the G-peak (excitation energy $\hbar\omega$ = 1.96 eV) and (c) for the D-peak. From panels (b) and (c), the two main effects, peak downshifts and decrease in peak intensities, are evident.

**Fig.2 –** Permanent increase of the low-field intensities of (a) the G-peak and (b) the D-peak. Data in panels (a) and (b) are recorded at the same run of Fig. 1b and 1c. (c) Dependence of the Raman G-peak intensities on the electric field ($E_{ext}$). – (d) Permanent decrease in sample conductance is measured and it does not recover after field release.

**Fig.3 –** (a) Dependence of the Raman shifts on the electric field ($E_{ext}$) as for eq.(6), assuming that the samples, initially at room temperature ($T_{min}$), heat proportionally to the electric field ($E_{ext}$). – Comparison with measured (b) G-peak and (c) D-peak frequencies. The lines represent data fits from eq.(6) with $T_{max}$ = 600K and $\tau$ as reported in the legend.

**Fig.4 –** Modifications of the RBMs by external field at (a) 1.96 eV (exciting both m- and s-SWNTs) and (b) 1.57 eV (exciting only s-SWNTs). Assignments of specific (n, m)-SWNTs are taken from Telg et al [31]. The *reversible* decrease for s-SWNTs (red) and the *reversible* increase for m-SWNTs (blue) are shown. In contrast, after field release, the *permanent* effects are always in increasing the RBM intensities for both s- and m-SWNTs at (c) 1.96 eV and (d) 1.57 eV.